\documentclass[pre, twocolumn, reprint,  superscriptaddress, eprintnumbers]{revtex4-2}
\usepackage[]{graphicx}
\usepackage{epstopdf}
\usepackage[dvipsnames]{xcolor}
\usepackage{bm}
\usepackage{hyperref}
\usepackage{framed,amsmath}
\usepackage{xcolor,colortbl}

\definecolor{Gray}{gray}{0.85}
\definecolor{LightCyan}{rgb}{0.88,1,1}

\newcolumntype{a}{>{\columncolor{Gray}}c}
\newcolumntype{b}{>{\columncolor{white}}c}

\begin{document}
	\counterwithin*{section}{part}
	\title{Dynamic hysteresis at a noisy saddle-node shows power-law scaling but nonuniversal exponent}
          
	\author{Satyaki Kundu}\email{kundusatyaki77@gmail.com}
	\affiliation{Indian Institute of Science Education and Research Kolkata, Mohanpur, Nadia 741246, West Bengal, India}
	\author{Ranjan Kumar Patel}
	\affiliation{Department of Physics, Indian Institute of Science, Bengaluru 560012, India}
	
	\author{Srimanta Middey}
	\affiliation{Department of Physics, Indian Institute of Science, Bengaluru 560012, India}

	\author{Bhavtosh Bansal}\email{bhavtosh@iiserkol.ac.in}
	\affiliation{Indian Institute of Science Education and Research Kolkata, Mohanpur, Nadia 741246, West Bengal, India}
	\date{\today}
	\begin{abstract}
		Dynamic hysteresis, viz., delay in switching of a bistable system on account of the finite sweep rate of the drive has been extensively studied in dynamical and thermodynamic systems. Dynamic hysteresis results from slowing of the response around a saddle-node bifurcation. As a consequence, the hysteresis area increases with the sweep rate. Mean-field theory, relevant for noise-free situations, predicts power law scaling with the area scaling exponent of $2/3$. We have experimentally investigated the dynamic hysteresis for a thermally-driven metal-insulator transition in a high quality NdNiO$_3$ thin film and found the scaling exponent to be about $1/3$,  far less than the mean field value. To understand this, we have numerically studied Langevin dynamics of the order parameter and found that noise, which can be thought to parallel finite temperature effects, influences the character of dynamic hysteresis by systematically lowering the dynamical exponent  to as small as $0.2$. The power law scaling character, on the other hand, is unaffected in the range of chosen parameters. This work rationalizes the ubiquitous power law scaling of the dynamic hysteresis as well as the wide variation in the scaling exponent between $0.66$ and $0.2$ observed in different systems over the last 30 years.
	\end{abstract}
	\maketitle
	\section{Introduction}
	Hysteresis, the history-dependent multivalued response from a system to an external drive, is a non-linear phenomenon frequently observed in physical \cite{Bertotti, Ikhouane}, electrical \cite{Ikhouane}, mechanical \cite{He_mech}, biological \cite{Noori}, ecological \cite{Raffaelli}, social \cite{Zahler} and economic \cite{Franz} systems. Although systematic studies of hysteresis date back to over a hundred years \cite{Ewing}, the rate dependence of hysteresis, viz.  dynamic hysteresis, was only discovered in 1986 \cite{Fidorra} and has been extensively studied since \cite{Bar_2018, cold_atom, binary_mixture,glass_wang,sbt_pan,pzt_liu,bpa_kim, He_fe, cu_jiang, Suen_fe, Bar_2021, Geng_2020, Rodriguez_2017,rao_1990, Zhong_1994, Lo_1990, Zheng_1998, Zhong_1995, Chakrabarti_rmp, Luse-Zangwill, resonator_Casteels,jung_1990, Dhar_1992, Somoza_1993, Zhang_solid_state, Pan_2003}. Dynamic hysteresis expresses the inability of a bistable system to keep up with the temporal change in the drive parameters.
	
	Many first-order thermodynamic phase transitions  exhibit dynamic hysteresis---the shift in transition points and the change in the area of the hysteresis loop show power law scaling with the rate of change of the drive parameter. This parameter  may be, for example,  the magnetic field \cite{zheng_2002,Zhong_2005,rao_1990,rao_1991,Luse-Zangwill, Sengupta-Marathe-Puri,zhong_new,Lo_1990,He_fe,cu_jiang,Suen_fe}, temperature \cite{Bar_2018, Bar_2021,Zheng_1998,rao_1991,Zhong_1995,cold_atom,binary_mixture,glass_wang}, or the electric field \cite{sbt_pan,pzt_liu,bpa_kim}. The area of the hysteresis loop scales as \cite{krapivsky}
	\begin{equation}
		A-A_0 \propto R^\gamma.
	\end{equation}
	Here $R$ can either be the frequency of the oscillatory drive or the sweep rate of a linearly changing parameter, $\gamma$ is the scaling exponent of the dynamic hysteresis, and $A_0$ is the area of the static hysteresis loop. The scaling exponent $\gamma$ is found to be non-universal and varying in a wide range \cite{Bar_2018, cold_atom, binary_mixture,glass_wang,sbt_pan,pzt_liu,bpa_kim, He_fe, cu_jiang, Suen_fe, Bar_2021, Geng_2020, Rodriguez_2017,rao_1990, Zhong_1994, Lo_1990, Zheng_1998, Zhong_1995, Chakrabarti_rmp, Luse-Zangwill, resonator_Casteels,jung_1990, Dhar_1992, Somoza_1993, Zhang_solid_state, Pan_2003}. The mean-field theory, valid for clean noiseless systems, predicts the exponent $\gamma$ to be 2/3 \cite{jung_1990, Luse-Zangwill, Bar_2018}. Disorder has been recently found to change the scaling exponent from the mean field value and yield $\gamma>$ 2/3 \cite{Bar_2021}.
	
	In this study, we have experimentally investigated the sweep-rate-dependent thermal hysteresis in NdNiO$_3$. NdNiO$_3$ is a well-studied example of electron correlation-driven hysteretic metal-insulator transition \cite{Ranjan_2020, Alsaqqa_2017, Sudipta_2021, Peil_2019, Middey_2016, Catalano_2018} accompanied by a symmetry-lowering bond disproportionation transition and magnetic transition.  We have found power law scaling of dynamic hysteresis with the power law exponent to be nearly $0.33$. This is far below the mean field value and cannot be understood by including disorder. We show that the dynamic scaling exponent and the area of the hysteresis loop decrease with the increase in noise, or equivalently,  thermal fluctuations.
	\subsection{Hysteresis in Landau Theory}
	Consider the mean field Landau free energy for a field-driven first-order phase transitions \cite{stauffer}
	\begin{equation}
		\label{eqn:free_exp_1}
		\mathcal{F}=\frac{1}{2}c_2(T-T_c)\phi^2+\frac{1}{4}c_4\phi^4-\mathcal{H}\phi.
	\end{equation}
	Here $c_2$ and  $c_4$ are constants, $T$ denotes the temperature, $T_c$ the critical temperature, $\mathcal{H}$ the external field,  and $\phi$ the scalar order parameter. Let us assume that $\phi$ is dimensionless. We can express Eq. \ref{eqn:free_exp_1} in dimensionless form by dividing it by some arbitrary energy $k_BT_0$,  that is,
	\begin{equation}
		\label{eqn:free_exp}
		F=\frac{1}{2}a_2\phi^2+\frac{1}{4}a_4\phi^4-H\phi.
	\end{equation}
	We have defined $F=\mathcal{F}/k_BT_0$, $a_2=c_2(T-T_c)/k_BT_0$, $a_4=c_4/k_BT_0$, and $H=\mathcal{H}/k_BT_0$.
	
	\begin{figure}[h!]
		\includegraphics[scale=0.24]{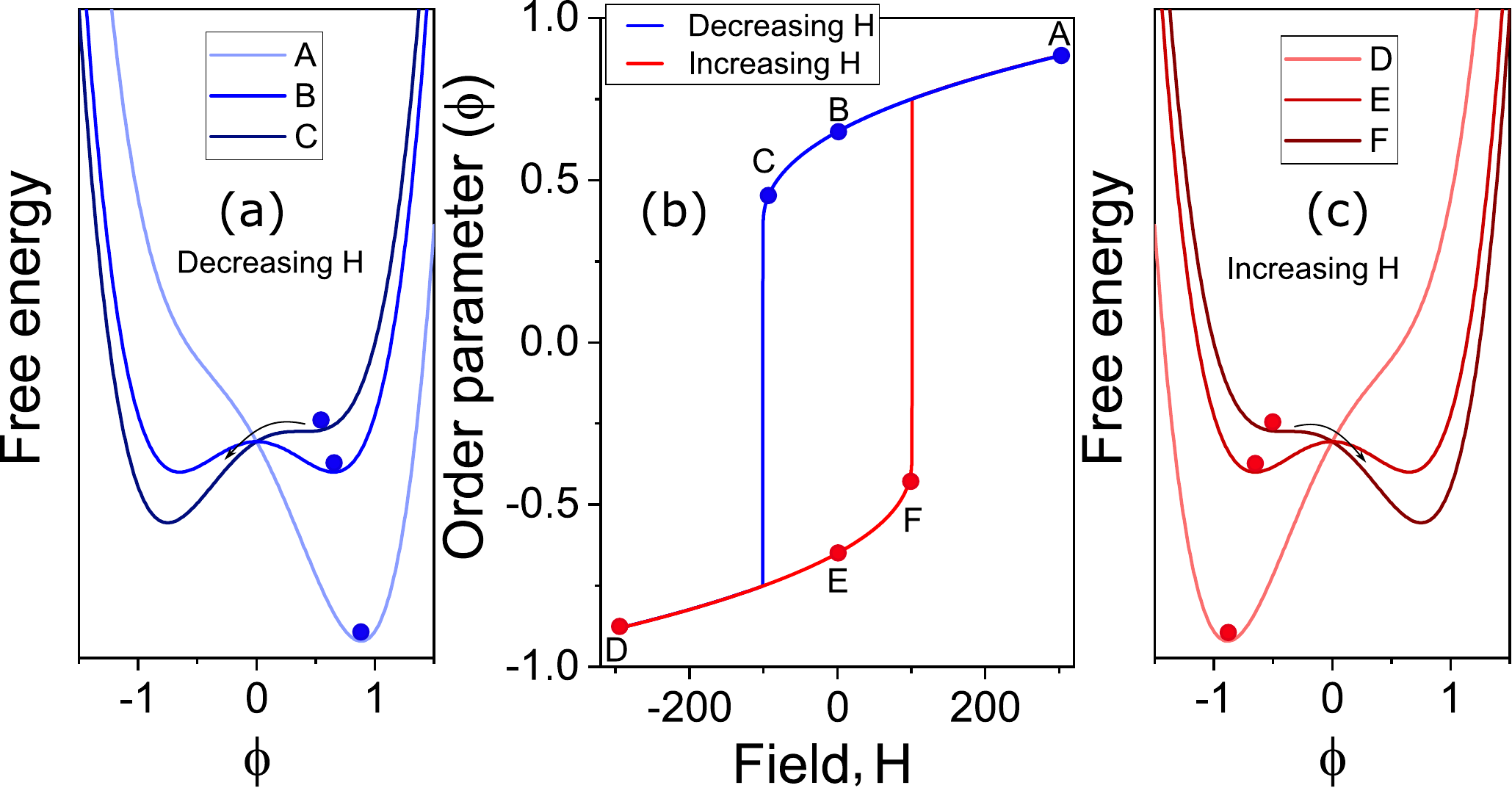}
		\centering
		\caption{\textbf{Origin of hysteresis in mean field theory}: (a) and (c) show the free energy (Eq. \ref{eqn:free_exp} with $a_2=-400$ and $a_4=948$) for different values of field $H$ during decreasing and increasing field respectively. B and E correspond to binodal point, where the two free energy minima have the same value. System depicted by solid ``$\circ$" persists to any minimum until spinodal point (C, F) when the nucleation barrier vanishes. After that the system rolls down to the other minimum. As a result (b) Order parameter shows hysteretic transition with field.}
		\label{Figure:potential_well}
	\end{figure}
	In the mean-field approximation, which ignores fluctuations, the Landau free energy (Eq. \ref{eqn:free_exp}) produces a first-order hysteretic phase transition with magnetic field when $a_2<$ 0 \cite{stauffer}. Hysteresis width increases with the increase of the absolute value of $a_2$ \cite{SM}.  The spinodal fields $H_s$ can be calculated in terms of $a_2$ and $a_4$ \cite{SM}, $H_s=\pm\sqrt{({4}/{27})({-a_2^{3}}/{a_4})}$.
	
	In Fig. \ref{Figure:potential_well}, we show an example of how hysteresis forms for a noise-free system in this model of the free energy with two competing minima. When the field decreases (increases) from some higher (lower) value A (D), the depth of global minimum (where the system is residing) decreases. At some value of the field (binodal), the depth of both minima becomes equal, depicted by points B and E. Further decreasing (increasing) field, we approach the spinodal point C (F) where the minimum vanishes. The region covered by the path $B \to C$ and $E \to F$ is metastable. Here the system is supersaturated because this region does not correspond to the global minimum of the free energy.
	
	To model this phenomenon quantitatively, we may simply construct an equation of motion assuming a fully-dissipative gradient dynamical system \cite{Gilmore, Strogatz_nld},  characterized by a spatially homogeneous nonconserved variable (the order parameter) $\phi (t)$ that now also has a time dependence,  viz.,
	\begin{equation}
		\label{eqn:TDL}
		{\partial\over \partial t} \phi=-\lambda{\delta F(\phi)\over \delta \phi}.
	\end{equation}
	The parameter $\lambda$ sets the time scale in the problem. Substituting the Landau free energy [Eq. \ref{eqn:free_exp}] in equation [Eq. \ref{eqn:TDL}],  we get
	\begin{equation}
		\label{eqn:bistable}
		\frac{d\phi}{d\tau}=A\phi-B\phi^3+H,
	\end{equation}
	where we have simplified the notation by defining $A=-\lambda a_2$, $B=\lambda a_4$ and $\tau=\lambda t$.
	
	This equation describes a dynamical system exhibiting saddle-node bifurcation \cite{Strogatz_nld} or a cusp catastrophe \cite{Gilmore}.
	In the absence of magnetic field $H$, there are two stable fixed points ($\phi=\sqrt{A/B}$ and $\phi=-\sqrt{A/B}$) and one unstable fixed point ($\phi=0$). The function $f(\phi)=A\phi-B\phi^3$ maximizes at $\phi=\sqrt{A/3B}$. $f(\sqrt{A/3B})=-(2A/3)\sqrt{A/3B}$. For finite $H<0$, the function $f(\phi)=A\phi-B\phi^3+H$ changes in such a way one stable fixed point and unstable fixed point start to come close. At the saddle-node bifurcation point P, where $H=-(2A/3)\sqrt{A/3B}$,  these two fixed points annihilate, resulting in an abrupt transition. Below $H=-(2A/3)\sqrt{A/3B}$, the curve has only one real solution. Similar behavior is also observed when $H$ is slowly increased from zero.
	
	\subsection{Origin of the 2/3 scaling exponent}
	\subsubsection{Perturbation theory}
\begin{figure}[h!]
		\includegraphics[scale=0.36]{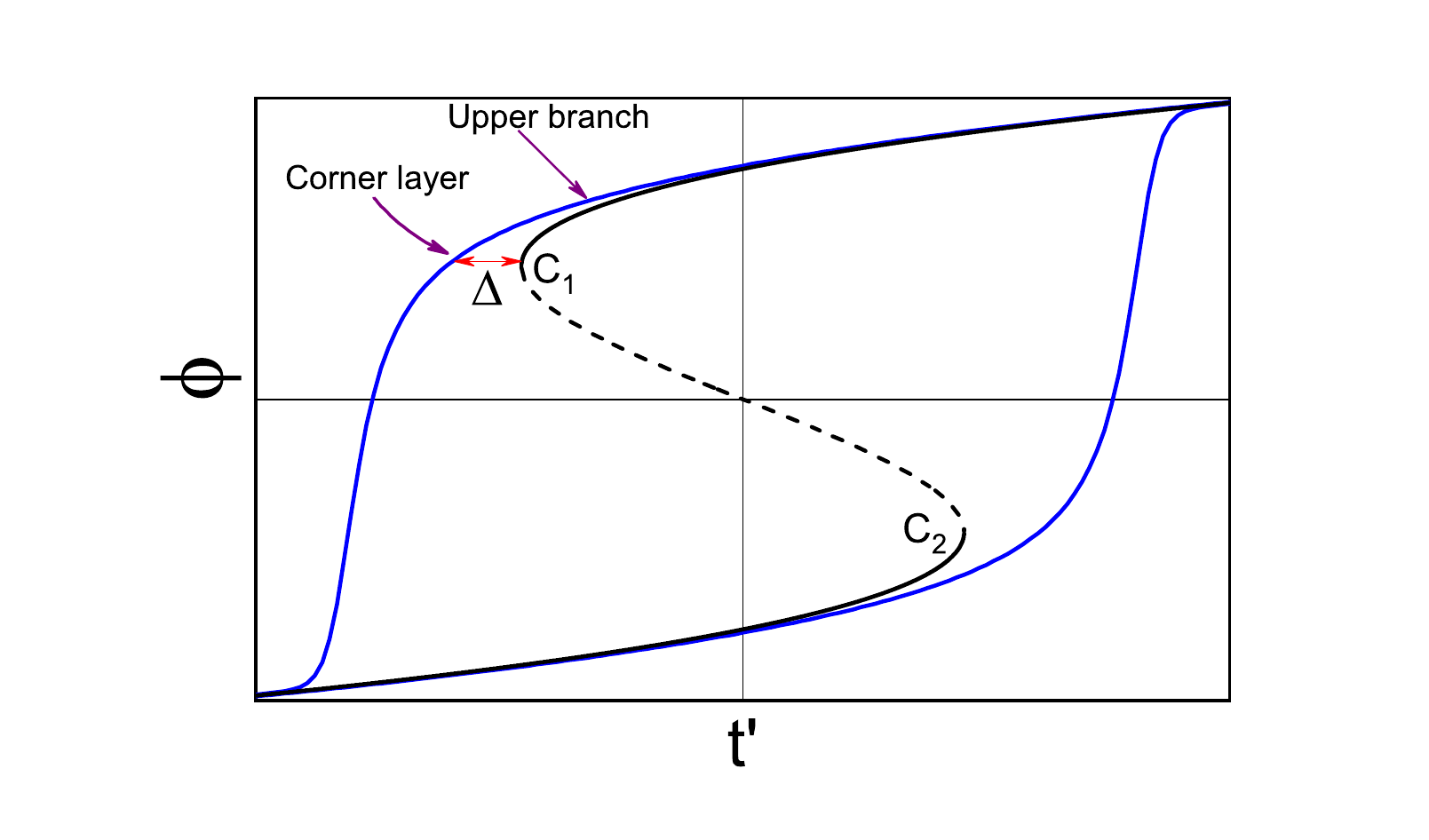}
		\centering
		\caption{For finite sweep rate, transition happens after a while compared to that of the static counterpart. The delay, $\Delta$ follows power law with sweep rate $\epsilon$. }
		\label{Figure:corner}
	\end{figure}

	The mean-field scaling exponent of the dynamic hysteresis can be estimated through an elegant perturbative approach \cite{krapivsky, holmes_perturb}. Assume that the magnetic field $H$ varies linearly with time, viz., $ H=\epsilon\tau$. We then have
	\begin{equation}
		\label{eqn:bistable_2}
		\frac{d\phi}{d\tau}=A\phi-B\phi^3+\epsilon \tau
	\end{equation}
	Let us define $t'=\epsilon\tau$, shift $\epsilon$ to the derivative term,
	\begin{equation}
		\label{eqn:bistable_3}
		\epsilon \frac{d\phi}{dt'}=A\phi-B\phi^3+t',
	\end{equation}
	and look for a perturbative solution (Lim $\epsilon \to 0$) by also expanding $\phi=\epsilon^0\phi_0+\epsilon^1 \phi_1+\epsilon^2 \phi_2... $ and matching powers of $\epsilon$. To order $\epsilon^0$ we have
	\begin{equation}
		\label{eqn:outer}
		A\phi_0-B{\phi_0}^3=-t'.
	\end{equation}
	The zeroth order solution, depicted in Fig. \ref{Figure:corner}, corresponds to the quasistatic limit and maps out the fixed points found earlier, but with the field replaced by a time variable $t^\prime$.
	Let us focus on the upper branch ($\phi_0 > 1$) of the curve in Fig. \ref{Figure:corner}. The end point under quasistatic drive, i.e., the spinodal, appears at  $(t^\prime_{sp},\phi_{sp})=(-\frac{2A}{3}\sqrt{\frac{A}{3B}},\sqrt{\frac{A}{3B}}$). This corner layer [$C_1$ in Fig. \ref{Figure:corner}] will be shifted by a delay $\Delta$ when the driving rate $\epsilon$ becomes non-negligible. We define corner variables by shifting the origin to the quasistatic corner point ($-\frac{2A}{3}\sqrt{\frac{A}{3B}},\sqrt{\frac{A}{3B}}$).
	\begin{equation}
		\label{eqn:corner}
		\begin{aligned}
			\tilde{t'} &= \frac{t'+\frac{2A}{3}\sqrt{\frac{A}{3B}}}{\epsilon^\gamma}\\
			\phi(t') &= \sqrt{\frac{A}{3B}}+\epsilon^\alpha \tilde{\phi_1}+\epsilon^{2\alpha} \tilde{\phi_2} + O(\epsilon^{3\alpha})
		\end{aligned}
	\end{equation}
	With respect to this new origin, we expect that the shift $\Delta\propto \epsilon^\gamma$, since any continuous function should be a power law close to the origin. Substituting in equation \ref{eqn:bistable_3}, we get
	\begin{equation}
		\epsilon^{1+\alpha-\gamma} \frac{d\tilde{\phi_1}}{d\tilde{t'}}=\epsilon^{2\alpha}\tilde{\phi_1}^2-\epsilon^\gamma\tilde{t'}+O(\epsilon^{3\alpha})
	\end{equation}
	Balance will be established if and only if
	\begin{equation}
		1+\alpha-\gamma=2\alpha=\gamma
	\end{equation}
	Thus $\gamma=\frac{2}{3}$ and $\alpha=\frac{1}{3}$.

	\begin{figure}[h!]
		\includegraphics[scale=0.4]{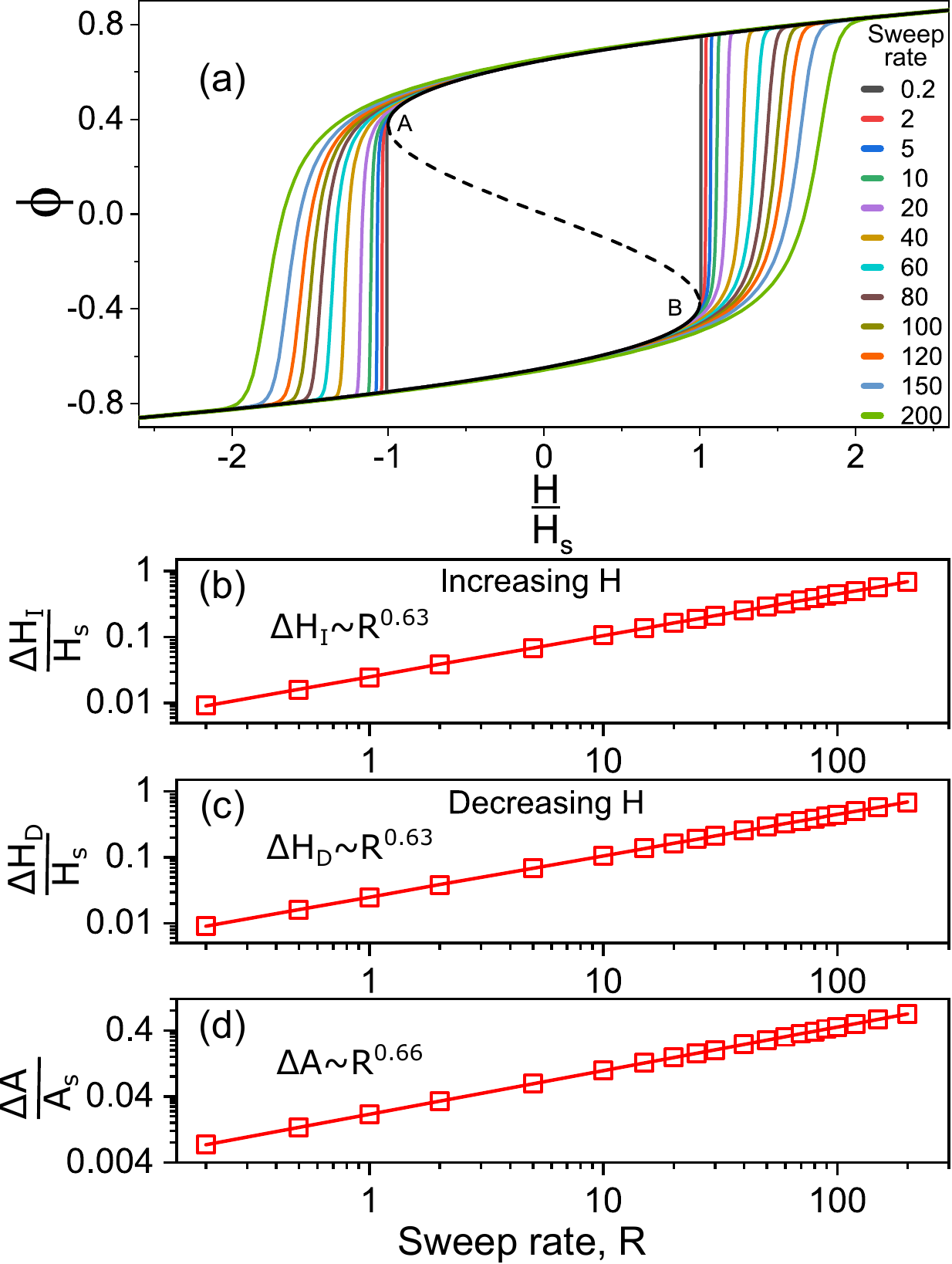}
		\centering
		\caption{\textbf{Dynamic hysteresis for noise free system}: (a) Transition temperatures along with the area under the hysteresis curve changes with sweep rate of driving field. Here we show how the hysteresis changes with 3 orders change in sweep rate. Solid black line depicts the bifurcation diagram of the system. A and B denote the boundary of the metastable phase i.e., spinodal. The curves between A and B indicate an unstable phase. (b) and (c) Shift in the transition field ($\Delta H_I$, $\Delta H_D$) with field sweep rate (R) during increasing and decreasing sweeping. (c) Increase of hysteresis area with the increase of field sweep rate has been plotted. The power laws fitting are depicted by solid lines. }
		\label{dyn_sim}
	\end{figure}
	\subsubsection{Numerical solution}
	$\gamma=\frac{2}{3}$ is also easily verified by numerically integrating Eq. \ref{eqn:TDL}. For concreteness, let $a_2=-400$, $a_4=948$, and $\lambda=0.0005$. Then the static spinodal field  $H_s = \pm 100$.  In the numerical solution shown in Fig. \ref{dyn_sim}(a), we linearly sweep the field  $-3\leq H/H_s\leq 3$ at rates varying between $0.2$ to $200$. Let us define $H_I$ and $H_D$, the transition fields for increasing and decreasing $H$ respectively, $A$ to be the area of ($\phi-H$) hysteresis loop. The dynamic scalings with the field sweep rate $R$ may be denoted as
	\begin{equation}
		\label{Equ:power_law}
		\begin{aligned}
			\Delta H_I &= H_I-H_{I0} &= k_{1}R^{\gamma_1},\\
			\Delta H_D &= H_D-H_{D0} &= k_{2}R^{\gamma_2},\\
			\Delta A &= A-A_0 &= k_{3}R^{\gamma_3},
		\end{aligned}
	\end{equation}
	where $k_1$, $k_2$, and $k_3$ are constants and $\gamma_1$, $\gamma_2$ and $\gamma_3$ are the power law scaling exponents. $H_{I0}$ and $H_{D0}$ are static spinodal field for increasing and decreasing $H$ respectively and $A_{0}$ is static hysteresis loop area. $H_{I0}$, $H_{D0}$ and $A_0$ must be nonzero for such transitions \cite{Bar_2018, Bar_2021, zheng_2002, Zhong_1995, rao_1990}.
	
	In Fig. \ref{dyn_sim} (b, c), we show the dynamic scaling of the spinodal fields $\Delta H_I$ and $\Delta H_D$ with the sweep rate $R$. The  scaling exponents $\gamma_1$ and $\gamma_2$ are both found to be $0.63$, close to the predicted value of $0.667$. Similarly in Fig. \ref{dyn_sim}(d), we show that the scaling exponent ($\gamma_3$) corresponding to increase of hysteresis loop area comes out to be $0.66$.
	\section{Experimental results}
	The motivation of this work comes from various temperature-driven first order phase transitions in solid state systems. Many such materials have a quasistatic hysteresis,  as well as a pronounced dynamic hysteresis even under a relatively slow temporal variation of temperature on scale of seconds \cite{Bar_2018}.

\begin{figure}[h!]
		\includegraphics[scale=0.25]{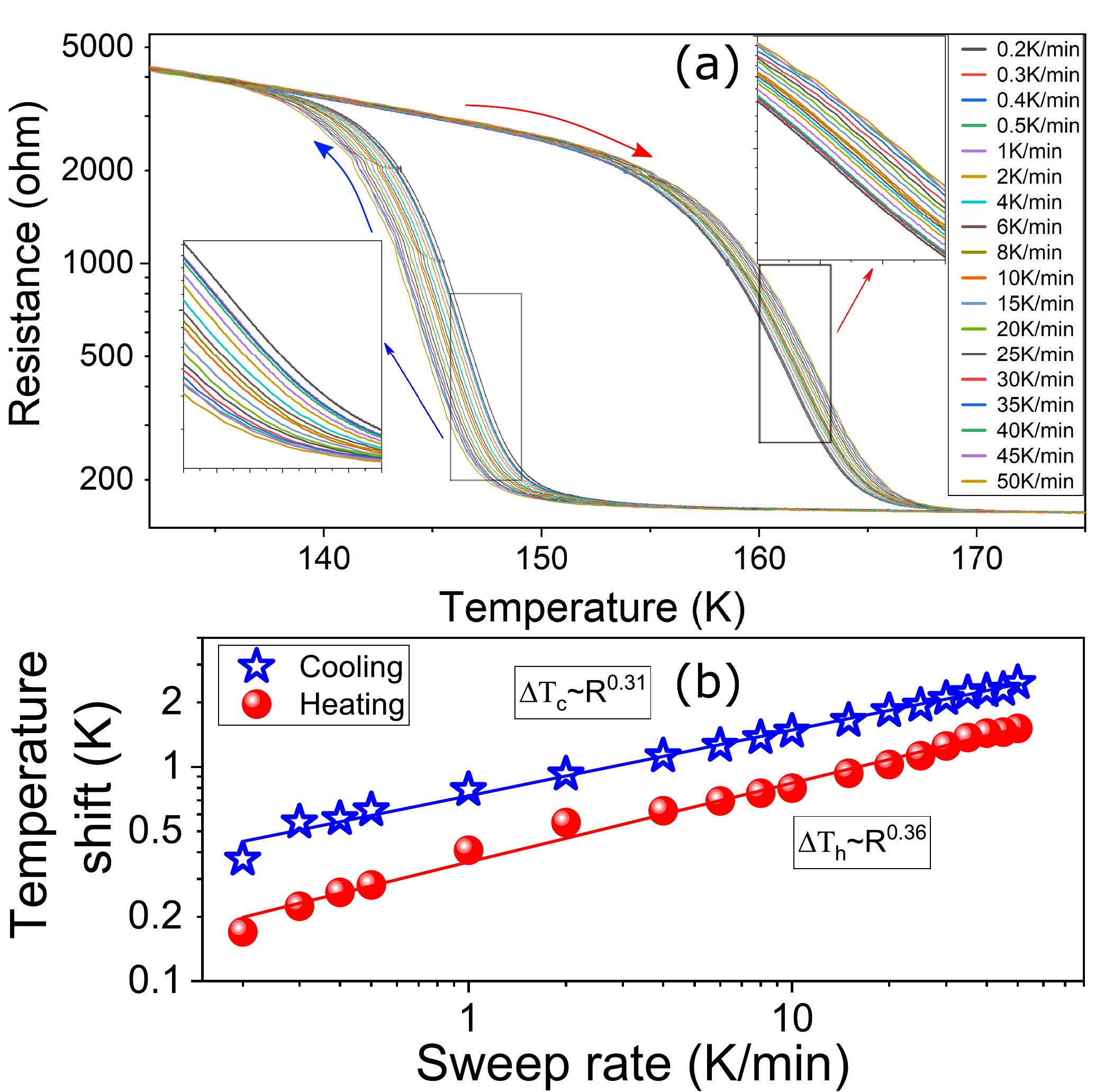}
		\centering
		\caption{\textbf{Dynamic hysteresis in NdNiO$_3$}: (a) Resistance of the NdNiO$_3$ sample shows thermal hysteresis. The heating and the cooling transition temperatures also shift with the temperature sweep rate. (b) Shift in heating and cooling transition temperature with sweep rate (R) of temperature ($\Delta T_c$ and $\Delta T_h$). $\Delta T_c$ and $\Delta T_h$ follow power law with sweep rate R with exponent 0.31 and 0.36 respectively. }
		\label{Figure:dyn_hs}
	\end{figure}

	Here we investigate the metal-insulator transition in very high quality single crystalline epitaxial thin film of NdNiO$_3$ (thickness: 15 unit cell $\sim$ 5.7 nm) grown on NdGaO$_3$ (110) substrate by pulsed laser deposition [see Ref. \cite{Ranjan_2020} for growth details and sample characterization] through resistance measurements under a linear temperature sweep.  Sample shows a phase transition from metallic to insulating phase around $145$ K during cooling and from insulating to metallic phase around $160$ K during heating. The sample resistance as a function of temperature, with the resistance plotted on the logarithmic scale, is shown Fig. \ref{Figure:dyn_hs}(a) for temperature sweep rates varying between $0.2$ K/min and 50 K/min.
	It is evident that the transition temperature shows temperature sweep rate dependent increase while heating and decrease while cooling, which also results in the overall increase of the area of the hysteresis loop.
	
	The thermal sweep rate dependent power-law shift in transition temperature, viz.,  $\Delta T_c= T-T_{0c} \propto R^{\gamma_c}$ and $\Delta T_h= T-T_{0h} \propto R^{\gamma_h}$  is shown in Fig. \ref{Figure:dyn_hs} (b). Here
	$T_{0c}$ and $T_{0h}$ are the quasistatic transition temperature under cooling and heating respectively, $\Delta T_c$ and $\Delta T_h$, the respective shifts. We find that the two exponents $\gamma_c= 0.31$ and  $\gamma_h=0.36$. The details of calculating the exponents are given in the Supplemental Material \cite{SM}. Both the scaling exponents are nearly $\frac{1}{3}$. Although similar values for the exponents have been previously observed in many theoretical \cite{rao_1990, Lo_1990, Zhong_1995, Chakrabarti_rmp} and experimental studies \cite{He_fe,pzt_liu}, these are very far from the mean field value of 2/3 observed in V$_2$O$_3$ \cite{Bar_2018}.
	
To connect the experimental observations to the formalism above (Eqs. 1--4), we can identify the temperature-dependent fraction of the insulating phase within the NdNiO$_3$ sample as a scalar non-conserved order parameter $\phi$. We can then convert the resistance of the sample to this insulator fraction using an effective-medium theory \cite{percolation}.
	\begin{equation}
		\label{Equ:percolation}
		\phi\frac{R_I^{-1/t}-R_E^{-1/t}}{R_I^{-1/t}+AR_E^{-1/t}}+(1-\phi)\frac{R_M^{-1/t}-R_E^{-1/t}}{R_M^{-1/t}+AR_E^{-1/t}}=0,
	\end{equation}
	where $R_I$ and $R_M$ are the respective resistances in the insulating and metallic phases. At a given temperature, resistance $R_E$ can be converted to insulator fraction $\phi$ using this relation. $A = (1-f_c)/ f_c$, $f_c$ being the volume fraction of metallic phases at the percolation threshold, and $t$ is a critical exponent which is close to 2 in three dimensions. The constant $f_c$ depends on the lattice dimensionality, and for 3D its value is $0.16$.
	
\begin{figure}[h!]
		\includegraphics[scale=0.17]{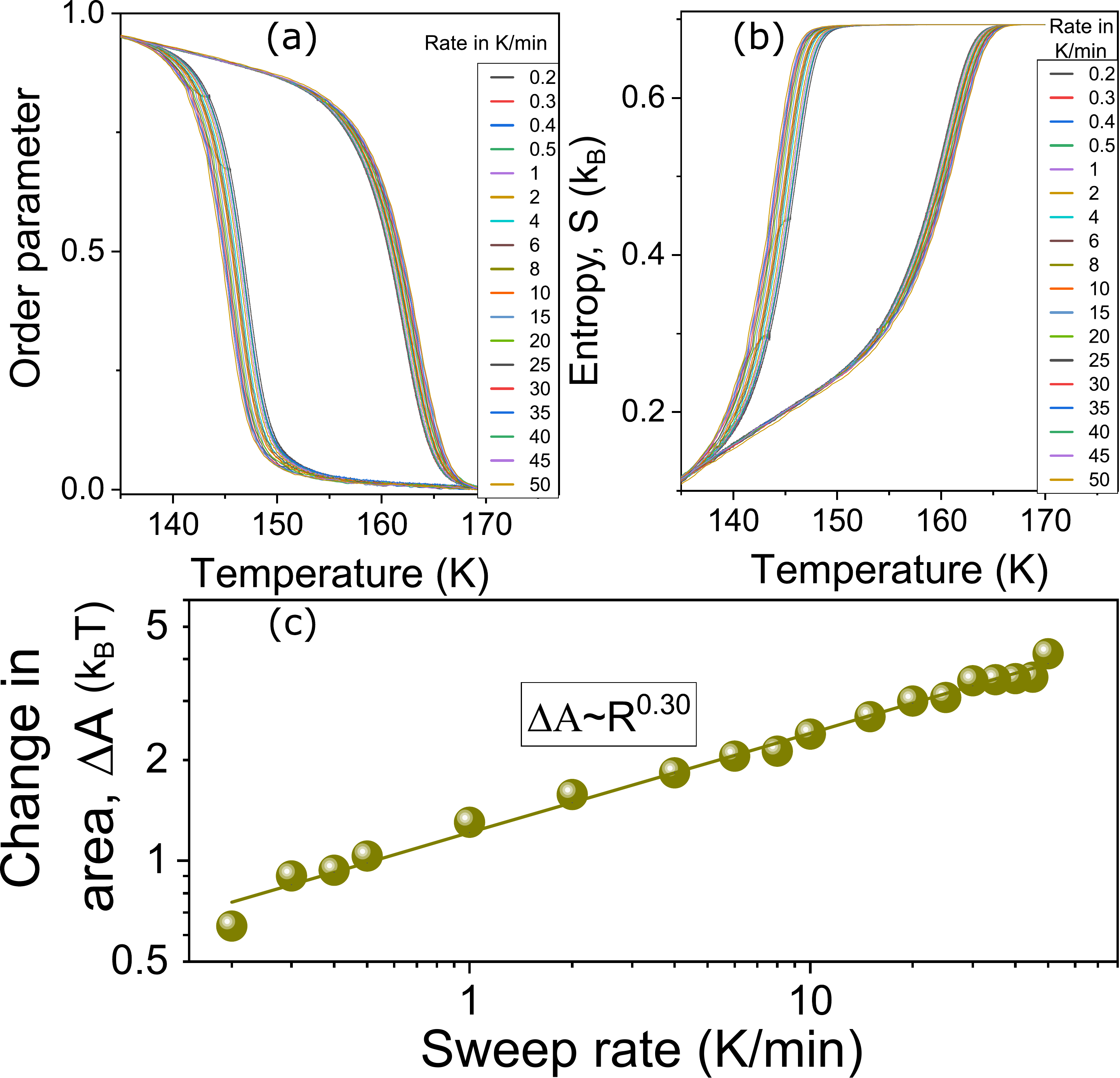}
		\centering
		\caption{\textbf{Estimation of scaling exponent from hysteresis area (Experimental)}: (a) Temporal evolution of the order parameter converted from resistance for different sweep rates. (b) Estimated entropy from the order parameter for different sweep rates (c) Increase in area of hysteresis loop of conjugate S-T plane with the increase of temperature sweep rate. Change in area, $\Delta A$ obeys $\Delta A \propto R^{0.3}$.}
		\label{Figure:entropy}
	\end{figure}

	In Fig. \ref{Figure:entropy} (a), we show that the hysteresis in the order parameter-temperature plane for different sweep rates follows a similar nature as resistance hysteresis curve. The order parameter is inferred from the experimentally measured resistance using Eq. \ref{Equ:percolation}. We can further estimate the entropy density of the system from this order parameter using Eq. \ref{Equ:ord_ent} \cite{Bar_2018, zheng_2002}. The area of the hysteresis loop in the conjugate coordinate, i.e., the entropy temperature (S--T) plane, indicates the energy loss or energy dissipation during the first-order phase transition
	\begin{equation}
		\label{Equ:ord_ent}
		S(\phi) =k_B \left[\ln2-\frac{1}{2} \left[ (1+\phi)\ln(1+\phi)+(1-\phi)\ln(1-\phi)\right]\right]
	\end{equation}
where $k_B$ is Boltzman constant. 

In Fig. \ref{Figure:entropy} (b), we show the evolution of the estimated entropy for different temperature sweep rates. Since temperature and entropy are conjugate variables, the area of hysteresis loop now has the proper interpretation of energy dissipation per cycle. The dynamic exponent for the area is found to be 0.3 [Fig. \ref{Figure:entropy}(c)]. This is very close to the exponents for the heating and the cooling temperature shifts. The scaling of the area of the hysteresis loop in the $T-S$ plane perhaps yields a better estimate of the scaling exponent because it is very hard to unambiguously associate the transition temperatures $T_{oc}$ and $T_{oh}$ in experimental systems where the transitions are never absolutely sharp. The inferred value of the scaling exponent is very sensitive to the choice of $T_{oc}$ and $T_{oh}$ (See Supplementary Material).
	
	\section{Langevin Dynamics}
	Let us now extend the calculation of Fig. \ref{dyn_sim} and study whether the dynamic scaling exponents are sensitive to noise and, indeed, if the power law scaling itself is preserved. Such noise can in a simple way account for thermal fluctuations which will invariably be present in any finite temperature measurement. We add a noise term $\zeta(t)$ in Eq. \ref{eqn:TDL} to get the Langevin equation
	\begin{equation}
		\label{eqn:TDL_noise}
		{\partial\over \partial t} \phi=-\lambda{\delta F(\phi)\over \delta\phi} + \zeta(t).
	\end{equation}	
	As usual, $\zeta(t)$ is assumed to be a $\delta$-correlated Gaussian random variable, viz., $\langle \zeta(t)\zeta(t^\prime)\rangle=\sigma^2 \delta(t-t^\prime)$,  with zero mean and variance $\sigma^2$. For the purpose of numerical solutions, the noise term $\zeta(t)$ is constructed as $\zeta(t)=\sigma N(0,1)/ \sqrt{t}$ \cite{Binney_1992, Romero_1992} where $N(0,1)$ is random number chosen from the standard Normal distribution. We define $s=\lambda t$ to make the Eq. \ref{eqn:TDL_noise} dimensionless, which in the discretized form now becomes
	\begin{equation}
		\label{eqn:TDL_3}
		\phi(s+\Delta s) = \phi(s)-{\delta F(\phi)\over \delta\phi} \Delta s + {\sigma N(0,1)\over \sqrt{\lambda}}\sqrt{\Delta s}.
	\end{equation}
	
	\begin{figure}[h!]
		\includegraphics[scale=0.25]{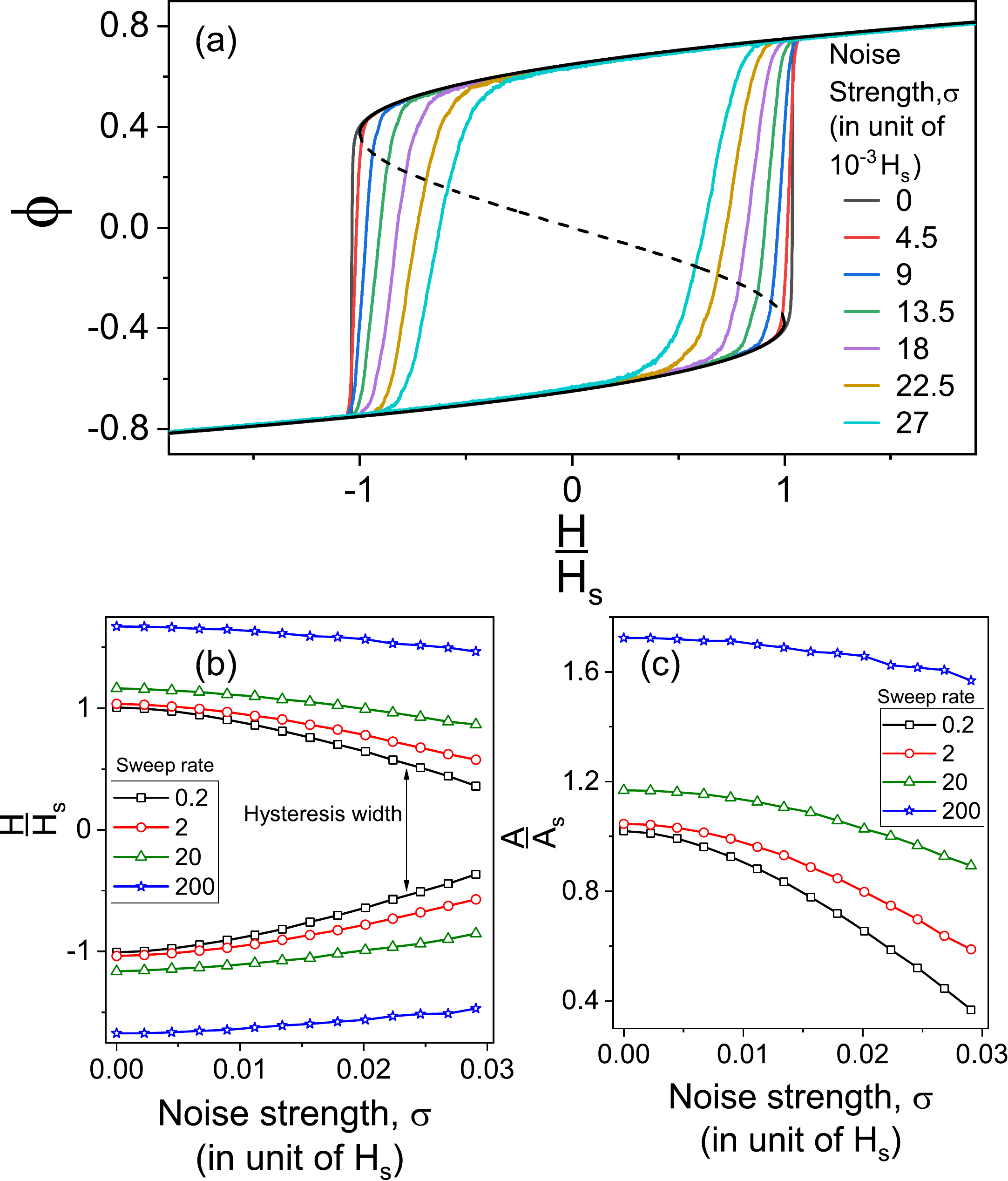}
		\centering
		\caption{\textbf{Shrinkage of hysteresis area with increase of noise}: (a) Black solid line is the bifurcation diagram of the system. Noise free system can exist in its metastable phase up to spinodals (A and B) when the nucleation barrier vanishes. Noise provides the activation energy to cross the nucleation barrier before the arrival of spinodal points. As a consequence transition occurs earlier and the area of the hysteresis loops shrinks with the increase of the noise strength. The noise strength in the legend is shown in the units of the spinodal field. We have also rescaled the transition field with spinodal field ($H_s$) and hysteretic area with the static hysteresis area of noise free system ($A_s$). (b) Transition fields versus noise strength for different sweep rates of field. With the increase in noise strength, the transition occurs closer to the origin (the binodal point) and thus the width of hysteresis shrinks. (c) Change in the area of the hysteresis loop with the noise strength for different field-sweep rates.}
		\label{Figure:hys_noise}
	\end{figure}
	The numerical results of the simulation of Eq. \ref{eqn:TDL_3} are summarized in Figs. 6--8 with the values of the parameters same as those used to generate Fig. 3. While a noise-free system would persist in its local minimum up to the spinodal point \cite{Hathcock_2021},  any finite noise decreases the depth of supersaturation by providing the activation energy to cross the nucleation before the spinodal is reached. As a result, the area of the hysteresis loop is expected to decrease with the noise strength \cite{Mahato_1994, Berglund_2006, Kuwahara_1995}. This is indeed seen in Fig. \ref{Figure:hys_noise} (a) where we plot the order parameter hysteresis for different noise strengths (scaled by spinodal field). In Fig. \ref{Figure:hys_noise} (b, c), we show how the hysteresis width and the area shrink with the noise strength for different field sweep rates.
	\begin{figure}[h!]
		\includegraphics[scale=0.3]{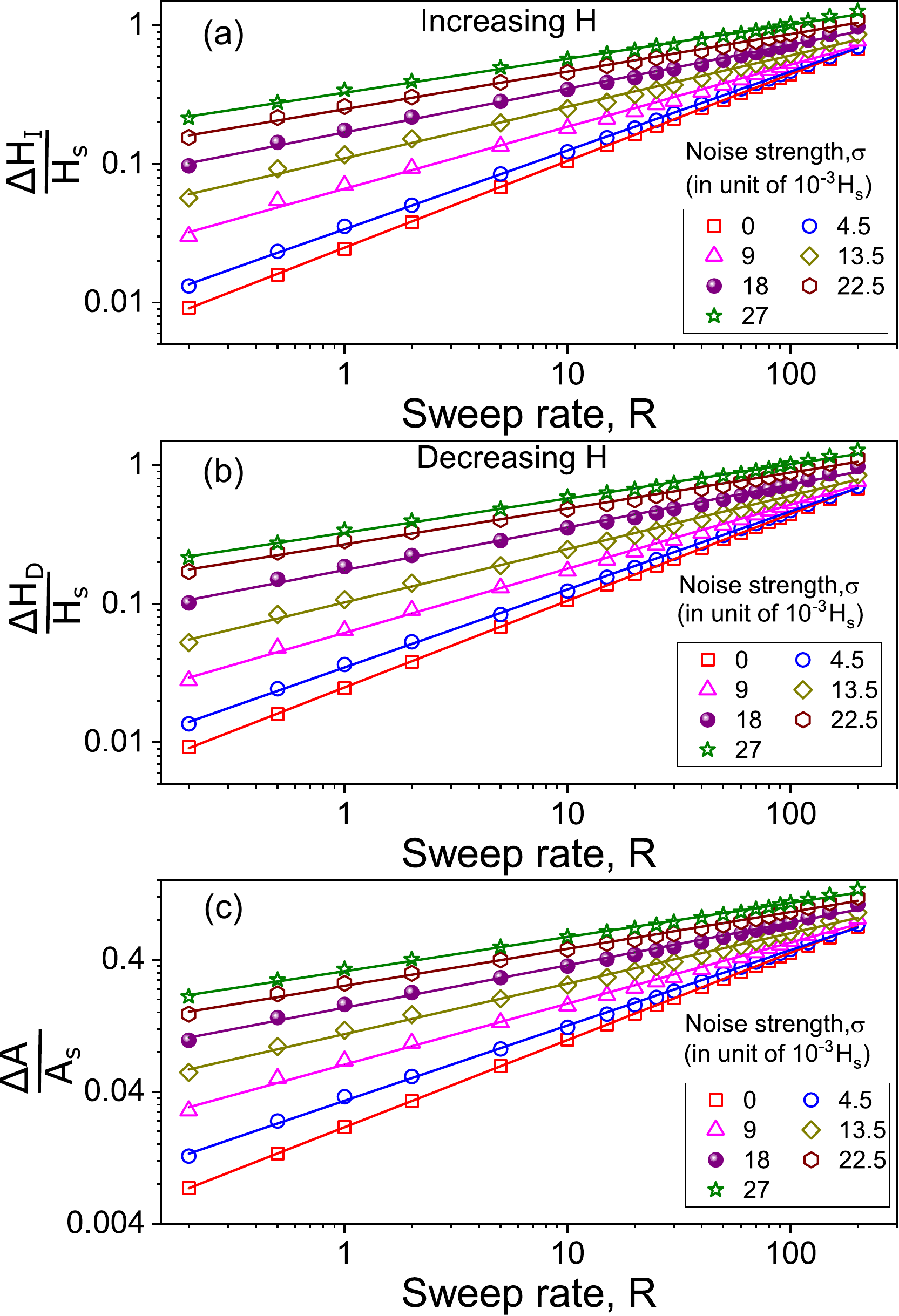}
		\centering
		\caption{\textbf{Decrease of the scaling exponent with the increase of noise}: (a) and (b) show the shift in the transition field ($\Delta H$) with field sweep rate ($R$) for different noise strengths for increasing and decreasing driving field, respectively. (c) Increase of the hysteresis loop area with the increase of the field sweep rate for different noise strengths. The power law fittings are depicted by solid lines.}
		\label{Figure:scaling_noise}
	\end{figure}
	These results are summarized in Fig. \ref{Figure:scaling_noise}, where we find that a power law scaling is still found over three orders of magnitude variation in the field sweep rate for different values of the noise strength both for the shift in the two transition fields [Fig. \ref{Figure:scaling_noise}(a, b)], as well as the change in the hysteresis loop area [Fig. \ref{Figure:scaling_noise}(c)]. But the scaling exponents are not constant and are found to monotonically decreases with the increase in noise. This is the main conclusion of the study. The change in the value as the function of the noise strength is plotted in Fig. \ref{Figure:scaling_sim}. The dynamic scaling exponent decreases to nearly $0.2$ for the noise strength $\sigma=0.03 H_s$. Our experimentally measured scaling exponent for NdNiO$_3$ matches the theoretical exponent for the effective noise strength $0.02 H_s$. Table \ref{tab:table1} further lists published values of $\gamma$  for a number of systems. Note that many of them fall in the rage of values observed in Fig. \ref{Figure:scaling_sim}.
	\begin{figure}[h!]
		\includegraphics[scale=0.35]{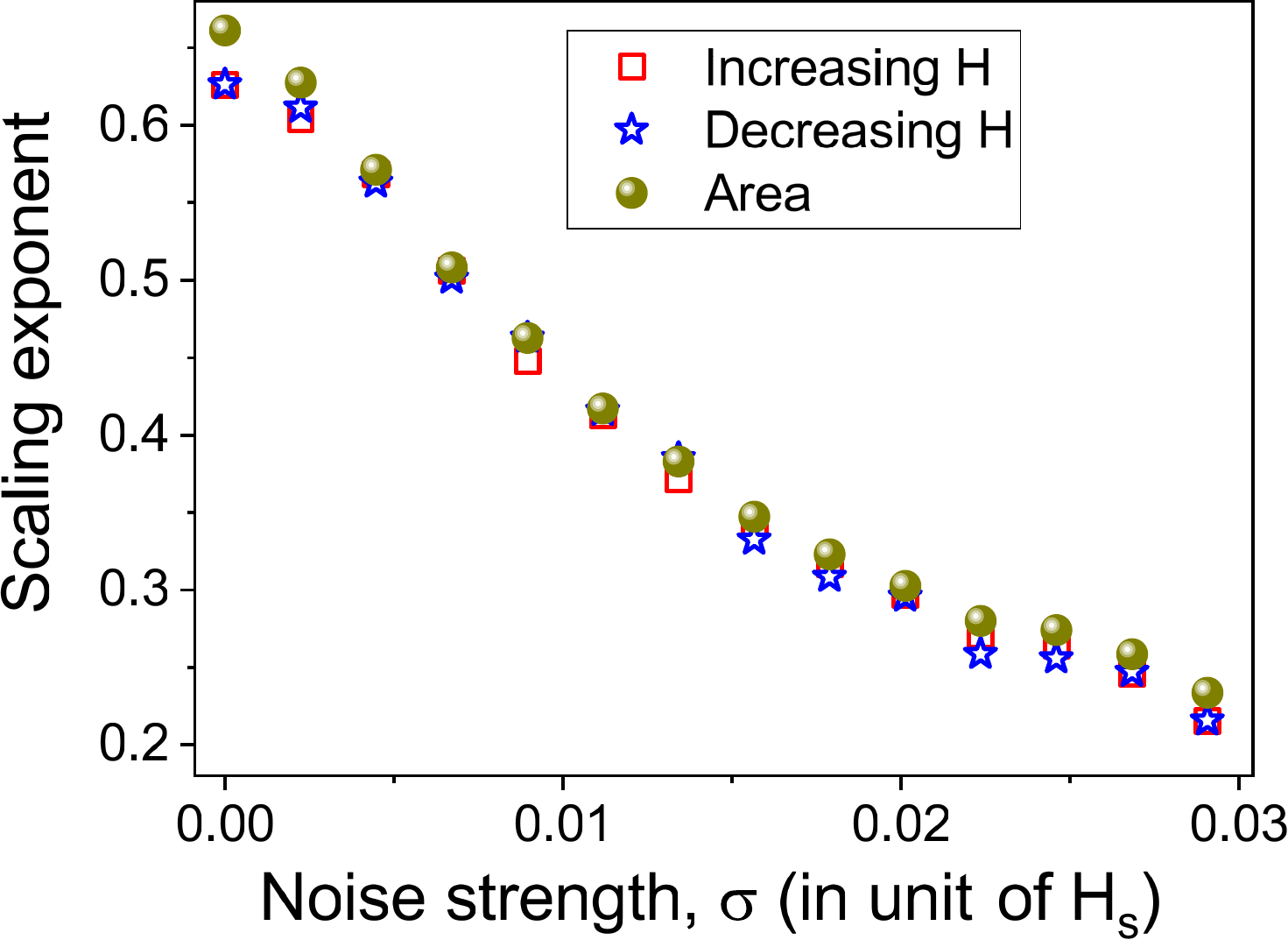}
		\centering
		\caption{Scaling exponent decreases from mean-field exponent with the increase of noise strength. Our experimentally obtained scaling exponent matches the theoretical exponent with noise strength 0.02.}
		\label{Figure:scaling_sim}
	\end{figure}
	\begin{table}[h]
		\caption{\label{tab:table1}%
			A table for dynamic scaling exponent for different types of system. (Adapted from ref. \cite{Bar_2022})
		}
		\begin{ruledtabular}
			\begin{tabular}{ l c }
					\textbf{Experiments} & \textbf{Scaling exponent}       \\
				\colrule
					PbTiO$_3$ \cite{Zhang_solid_state}                                                   & $1$                             \\
				FeMn alloy \cite{Pan_2003}                                                                         & $1$                             \\
				NdNiO$_3$ thin film (disorder) \cite{Prajapati_2022}                                               & $0.94$ (Heat) $0.98$ (Cool)     \\
				Heusler alloy \cite{Bar_2021}                                                                      & $0.93$ (Heat) $0.85$ (Cool)     \\
				Glass-forming glycerol \cite{glass_wang}                                                           & $0.88 \pm 0.09$                 \\
				Binary Mixture \cite{binary_mixture}                                                               & $0.692$                         \\
				\rowcolor{pink}
					Co/Cu film \cite{cu_jiang}                                                    & $0.66$                          \\
				\rowcolor{pink}
					SBT thin films \cite{sbt_pan}                                                 & $0.66$                          \\
				\rowcolor{pink}
					Cold atom (Mean field) \cite{cold_atom}                                       & $ 0.64 \pm 0.04 $               \\
				\rowcolor{pink}
					$V_2O_3$ \cite{Bar_2018}                                                      & $ 0.62 $ (Heat) $ 0.64 $ (Cool) \\
				\rowcolor{pink}
					BPA bulk system \cite{bpa_kim}                                                & $0.40$                          \\
				\rowcolor{SkyBlue}
					NdNiO$_3$ thin film (high quality)                                         & $0.36$ (Heat) $0.31$ (Cool)     \\ 
				\rowcolor{pink}
					PZT thin films \cite{pzt_liu}                                                 & $0.33$                          \\
				\rowcolor{pink}
					Fe/Au film \cite{He_fe}                                                       & $0.31$                          \\ 
				Fe/W film \cite{Suen_fe}                                                                           & $0.02$                          \\
				Optical cavity \cite{Geng_2020,Rodriguez_2017}                                                     & $-1$                            \\ \hline
				\textbf{Numerical simulations}                                                                     &                                 \\ \hline
				$(\Phi^2)^3$ model \cite{Zhong_1994}                                                               & $0.7$                           \\
				Four-spin Ising (FCC) (MF) \cite{Zheng_1998}                                                       & $0.7 \pm 0.05$                  \\
				Quantum resonator (MF) \cite{resonator_Casteels}                                                   & $0.66$                          \\
				Mean-Field \cite{Luse-Zangwill}                                                                    & $0.66$                          \\
				$(\Phi^2)^2$ model \cite{rao_1990}                                                                 & $ 0.33$                         \\
				$(\Phi^2)^2$ model \cite{Zhong_1994}                                                               & $0.5$                           \\
				Four-spin Ising (SC) (MF) \cite{Zheng_1998}                                                        & $0.47 \pm 0.05$                 \\
				Ising 3D Monte-carlo \cite{Chakrabarti_rmp}                                                        & $0.45$                          \\
				Ising 2D Monte-carlo \cite{Lo_1990, Zhong_1995, Chakrabarti_rmp}                                   & $ 0.36$                         \\ 
					\textbf{Analytical arguments}                                  &                                 \\ \hline
				Mean field \cite{jung_1990}                                                                        & $0.66$                          \\
				$(\Phi^2)^2$ model \cite{Dhar_1992, Somoza_1993, Zhong_1995}                                       & $0.5$                           \\ 
			\end{tabular}
			\end{ruledtabular}
		\end{table}
		\section{Discussion}
		Apart from the experiments on NdNiO$_3$ reported here, the motivation for this study also comes from the continued interest in dynamic hysteresis over the past three decades \cite{Bar_2018, cold_atom, binary_mixture,glass_wang,sbt_pan,pzt_liu,bpa_kim, He_fe, cu_jiang, Suen_fe, Bar_2021, Geng_2020, Rodriguez_2017,rao_1990, Zhong_1994, Lo_1990, Zheng_1998, Zhong_1995, Chakrabarti_rmp, Luse-Zangwill, resonator_Casteels,jung_1990, Dhar_1992, Somoza_1993, Zhang_solid_state, Pan_2003}.
		Dynamic hysteresis is a manifestation of the critical slowing down at the saddle-node bifurcation point \cite{Tredicce_2004,Scheffer-Review}. In thermodynamic contexts, this saddle-node is identified with the spinodal point. Note that the spinodal point $H=H_s$ is analogous to the critical point because of the diverging susceptibility, viz.,
		$$\chi^{-1}\equiv {\partial^2 F\over\partial \phi^2}|_{_{H=H_s}}\to 0, $$
		Therefore from Eq. \ref{eqn:TDL}, the response time of the system should also diverge \cite{Kundu_2020, Scheffer-Review}. This makes the case for the dynamic scaling exponent $\gamma$ to be treated at par with the other critical exponents \cite{Zhong_2005}.
		
		But for systems with disorder and/or noise, the nature of the spinodal singularity is not very well understood \cite{Nandi_2016, Kundu_2020}. Firstly, in the numerous studies on dynamic hysteresis, the value of $\gamma$ has been found anywhere between $1$ and $0.3$  [see Table \ref{tab:table1}] and therefore, unlike the critical exponents of a continuous transition, it does not seem universal.  Sometimes $\gamma$ also differs from sample to sample for the same material. This includes NdNiO$_3$ where $\gamma\approx 0.95$ was recently measured \cite{Prajapati_2022} as opposed to $\gamma\approx 0.3$ measured in this work. Value of $\gamma>2/3$  are generally seen for  heavily disordered systems, for example glass-forming glycerol \cite{glass_wang}, polycrystalline Heusler alloy \cite{Bar_2021}, and also in simulations of the zero-temperature random-field Ising model \cite{Bar_2021, Bar_2022}. The variation of the scaling exponent can perhaps be reconciled by Harris criterion-like arguments for the spinodal singularity \cite{Nandi_2016, Liu-klein}.
		
		$\gamma<2/3$ is also very often seen both experimentally and in simulations. It was argued very early on that thermal fluctuations must yield corrections to the mean field result \cite{He_fe}. But we note that the spinodals are strictly defined only for noise-free (zero temperature or mean field) systems. Any finite temperature should mask the spinodal singularity and make it physically inaccessible \cite{binder_rpp, Klein_fluctuations}.  For example, it was recently shown that the barrier escape times (which would diverge at the noiseless bifurcation point) lose all signatures of the spinodal singularity even in the presence of infinitesimal noise \cite{Hathcock_2021}.
		
Let us consider the experimental observations now for finite-temperature thermodynamic systems, like NdNiO$_3$ studied in this paper and many others listed  in Table \ref{tab:table1}. The fact that these are found to have non-zero hysteresis  is surprising, given the extreme sensitivity of the spinodal to noise. Somehow, for such solid state phase-change materials, there is a strong suppression of fluctuations that---at least over the laboratory time scales of days or months (viz., more 15 orders of magnitude larger than the phonon time scales)---the system can persist in the supersaturated metastable phase. Since this is what is also predicted by the mean field theory, it has been tempting (on purely empirical grounds) to associate the corner layer of Fig. \ref{Figure:corner} with the mean-field spinodal, even for finite temperature situations and indeed signatures of singularity at the end points of the metastable phase are experimentally found \cite{Bar_2018, Kundu_2020}. The reason for this mean-field like behaviour is generally attributed to long-range strain fields \cite{Klein_fluctuations, Klein_Nucleation_elastic}---recall that the infinite-range Ising model is equivalent to a nearest-neighbor Ising model  in the mean field approximation \cite{stauffer}. With infinite-range interactions, different parts of the system are perfectly coupled and the effect of local stochastic forces averages out to zero. Consequently, even at finite temperatures the system is effectively fluctuation-free.  

It is empirically seen that long- (but finite) range interactions preserve the key aspects of the mean field physics---finite hysteresis widths under quasistatic drive and the power-law scaling of the transition temperature and hysteresis loop area with the temperature sweep rate. But, as is seen in Table 1, the scaling exponents themselves become non-universal, dependent on the specifics of the material system.

In this work, we have argued that this departure from the ideal mean-field (infinite-range/zero-temperature) behavior in finite-temperature systems with finite-ranged interactions can be reproduced by adding small noise in the order parameter dynamics [Eq. \ref{eqn:TDL_noise}]. The dynamic hysteresis scaling exponent shows a monotonic decrease as a function of the noise strength. While this noise represents thermal fluctuations, the long-range interactions ensure that only a small fraction (a few long wavelength Fourier modes) of the total thermal noise influences the dynamics. The exact value of this fraction would be dependent on the microscopic details of the system under consideration and would vary across systems. Hence it was used as a parameter in the simulations.

\section{Conclusions}
We have experimentally studied the dynamic scaling of thermal hysteresis around the Mott transition in NdNiO$_3$. We find a power law scaling for the hysteresis loops but the scaling exponent is non-universal. For the high quality film sample studied here, the exponent is much lower than the mean field value of $2/3$. To reconcile this result in particular, and the large variation in the scaling exponent found across systems (Table \ref{tab:table1}) in general, we studied the problem via fully dissipative Langevin dynamics of the order parameter in a model for a field-dependent first-order phase transition. The dynamic scaling exponent in this minimal model is strongly renormalized by the noise strength to values as small as $0.2$, which is indeed much less than the mean field prediction $\gamma=2/3$. Surprisingly, the feature of the power law scaling itself is robust to noise.   
		
		These numerical results show that even in the presence of small noise (a few percent of the spinodal field), there is significant portion of the hysteretic metastable phase where the nucleation barrier is large enough to allow for a regime of parameter sweep rates for the approach to the bifurcation point where a finite hysteresis loop area will be measured. That there is a power law scaling (over at least three orders of magnitude of the sweep rate (Fig. 7) is surprising. It indicates that there is a clear separation of time scales. The reason for this is that the noise-induced activated barrier crossing time is exponentially large in the nucleation barrier height. These results point to the robustness of the spinodal singularity, at least in this one specific dynamical setting. As Table 1 demonstrates, the results presented here in the context of a dynamic hysteresis for noisy saddle node bifurcations have implications for thermodynamic systems undergoing first-order phase transitions and connect up with a larger problem of the spinodals in a real finite temperature system \cite{Kundu_2020, Bar_2018, Klein_fluctuations, Mori-Miyashita-Rikvold, Miyashita-Konishi-Nishino-Tokoro-Rikvold}.
		\section{Acknowledgement}
		It is pleasure to thank Pradeep K. Mohanty for fruitful advice  and Tapas Bar for discussions. SK thanks Council of Scientific and Industrial Research (CSIR), India for financial support. SM acknowledges SERB, India  (I.R.H.P.A Grant No. IPA/2020/000034) and MHRD, Government of India under STARS research funding (STARS/APR2019/PS/156/FS) for financial support. BB thanks the Science and Engineering Research Board (SERB), Department of Science and Technology, Government of India, for the Core Research Grant (No. CRG/2018/003282 and CRG/2022/008662).

	     \setcounter{section}{0}
		\begin{center}
			{\large {\bf Supplemental Material}}
		\end{center}
		\noindent
		In the Supplementary Material we discuss the Landau theory of field-driven first-order transitions in the language of catastrophe theory and the method of calculating scaling exponent from numerical and experimental data.

			\section{Field driven first order phase transition}
		\subsection{Introduction}
		
		The Landau free energy for field driven FOPT can be represented as [also described in Eq. 3 (main text)]
		\begin{equation}
			\label{eqn:free_exp}
			F=\frac{1}{2}a_2\phi^2+\frac{1}{4}a_4\phi^4-H\phi.
		\end{equation}
		
		The dynamical equation which follows by inserting the free energy Eq. 4 (main text) is equivalent to the standard unfolding of cusp catastrophe \cite{Gilmore} and $a_2=0$ is the cusp point. Below the cusp point ($a_2<0$), noise free system shows hysteresis of  order parameter with field. The width of the hysteresis loop increases with the increase in the absolute value of $a_2$. In Fig. \ref{Figure:width_a}, we show how the width of the order parameter hysteresis loop shrinks with the increase of $a_2$ from some negative value. When $a_2\ge 0$, there is no hysteresis. The dotted line represents the static (unstable) solution of the free enrgy which is always inaccessible in dynamic systems. 
		
		\renewcommand\thefigure{1 (supplement)}
		\begin{figure}[h!]
			\includegraphics[scale=0.33]{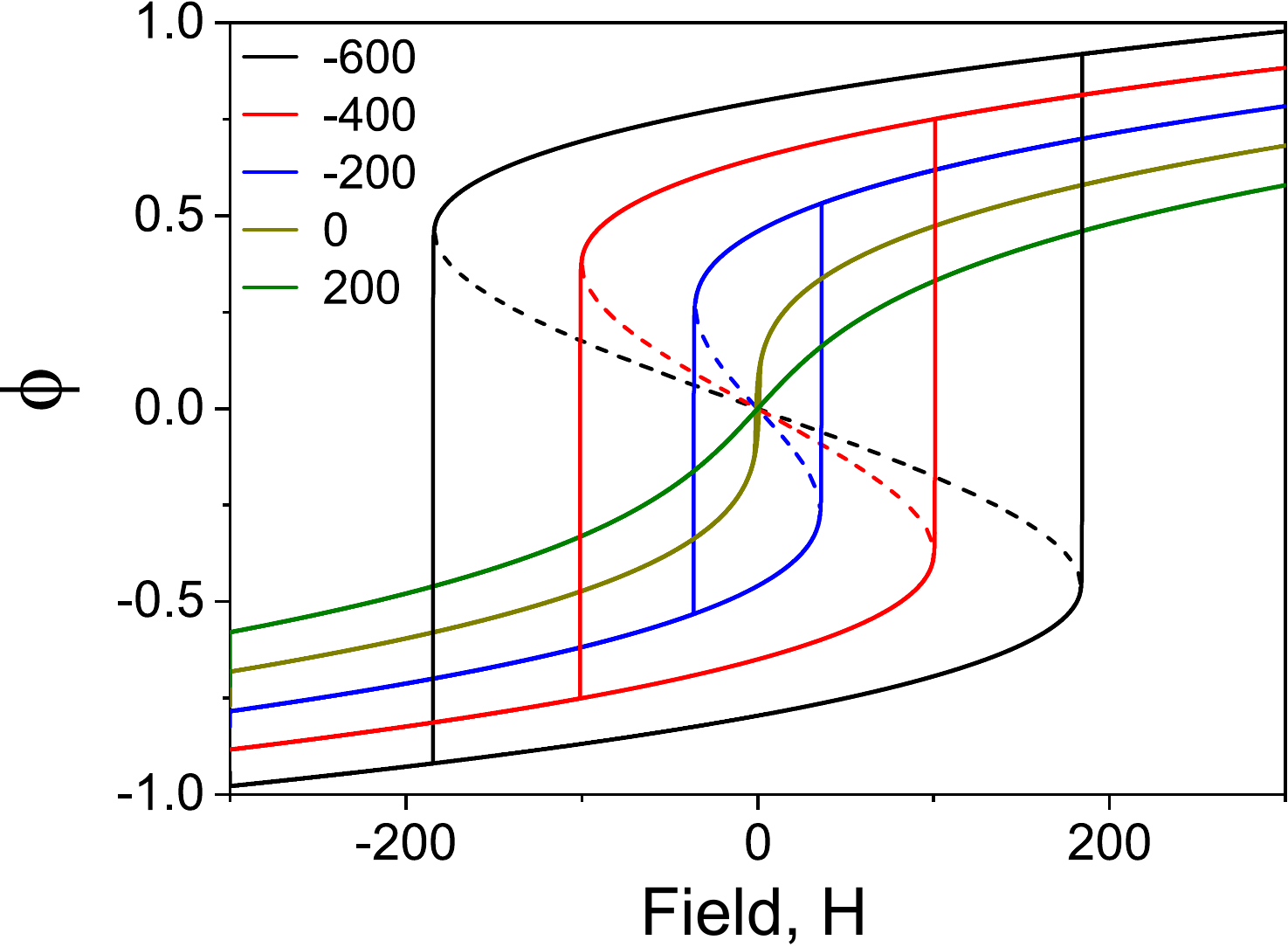}
			\centering
			\caption{Width of the hysteresis loop shrinks with the increase of $a_2$ and vanishes $a_2=0$. When ($a_2>0$) transition happens continuously (second order) from one phase to other phase. Legend represents the value of $a_2$. In our model we choose $a_4=948$. Dotted line represents the unstable solution.}
			\label{Figure:width_a}
		\end{figure}
		
		\renewcommand\thefigure{2 (supplement)}
		\begin{figure}[h!]
			\includegraphics[scale=0.35]{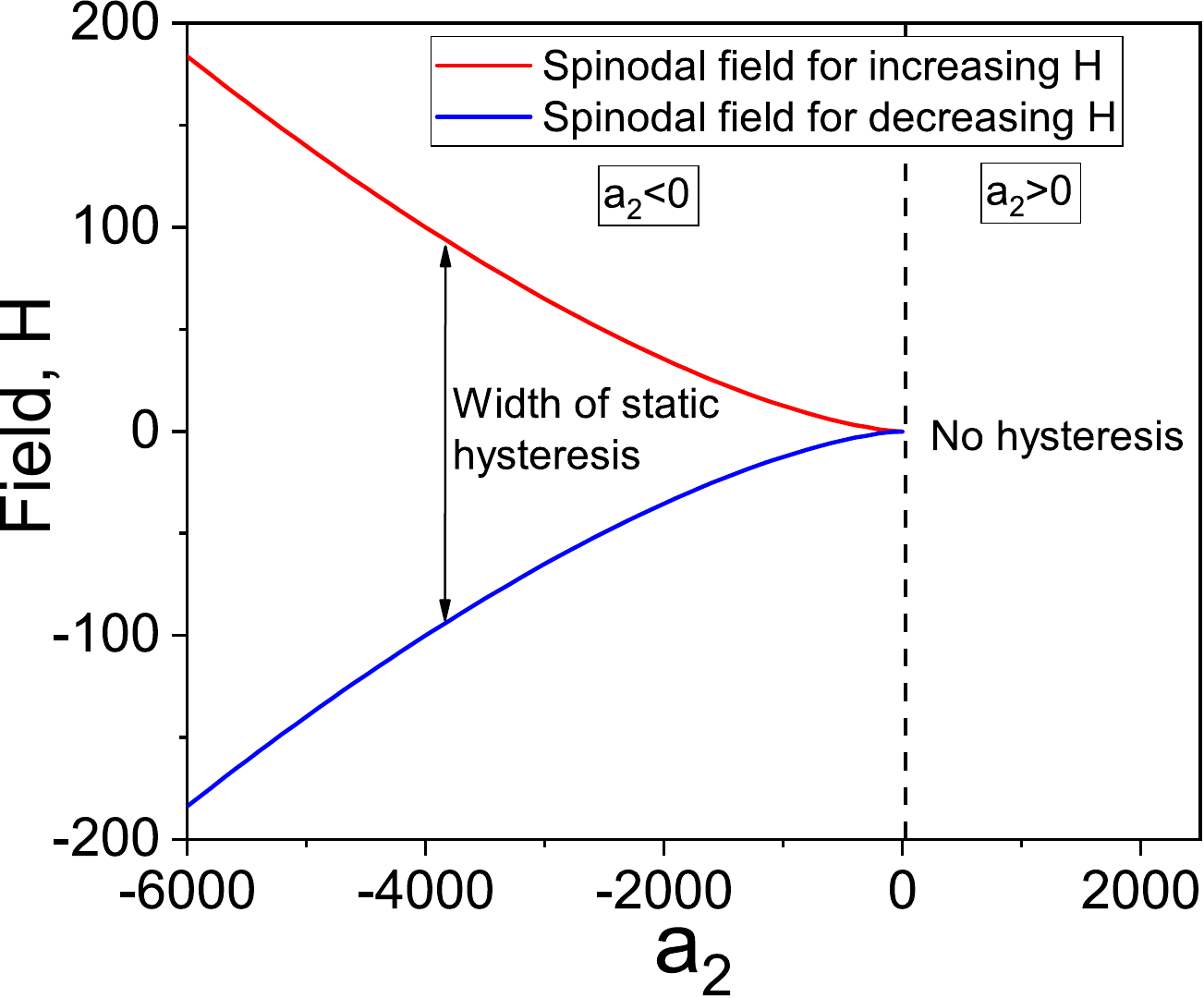}
			\centering
			\caption{($a_2=0$) is the boarder line (black dashed line) between first order hysteretic phase transition and second order transition. The absolute value of spinodal field increases with the decrease of temperature from critical temperature. As a result width of hysteresis increases for lower temperature. Cooling and heating spinodal field ($H_s$) is depicted with blue and red lines which follow Eq. \ref{eqn:spinodal_field}. We choose $a_2=-400$ for performing all simulation. Spinodal field, $H_s(a_2=-400)= \pm 100$.} 
			\label{Figure:bif}
		\end{figure}
		
		\subsection{Estimation for Spinodal field}
		Since the free energy is a biquadratic function (Eq. \ref{eqn:free_exp}) of order parameter, there exist two minima and a maximum under certain circumstances. We can find all the extrema ($\phi_s$) of the free energy by considering the fact that first derivative of free energy with respect to order parameter vanishes at all extrema i.e., $\frac{\partial F}{\partial \phi}|_{\phi=\phi_s}  =0$
		
		\begin{equation}
			a_2\phi_{s}+a_4\phi_{s}^3-H=0
		\end{equation}
		
		Solving the cubic equation we get 
		\begin{equation}
			\phi_{s}=\sqrt[3] {\frac{H}{2a_4} \pm \sqrt{\frac{H^2}{4a_4^2}+\frac{a_2^{3}}{27a_4^3}}} - \frac{\frac{a_2}{a_4}}{3 \sqrt[3] {\frac{H}{2a_4} \pm \sqrt{\frac{H^2}{4a_4^2}+\frac{a_2^{3}}{27a_4^3}}}}
		\end{equation}
		
		If a system is free from thermal fluctuations, the metastable state of the system can persist up to the limit of stability, called spinodal point, where the nucleation barrier disappears. In other words, phase evolution of the system at the spinodal is a roll-down to the global minimum, due to the vanishing of the barrier. After crossing spinodal, the free energy associated with the system has only one minimum. At the spinodals ($H=H_s$)
		
		\begin{equation}
			\frac{H_s^2}{4a_4^2}  + \frac{a_2^{3}}{27a_4^3}=0
		\end{equation}
		
		\begin{equation}
			H_s=\sqrt{-\frac{4}{27}\frac{a_2^{3}}{a_4}}
			\label{eqn:spinodal_field}
		\end{equation}
		
		In Fig. \ref{Figure:bif}, we show the imperfect bifurcation diagram for this model. In the regime $a_2>0$, there is no hysteresis, as also seen in Fig. \ref{Figure:width_a}. In the regime $a_2<0$, hysteresis width increases as $a_2$ decreases.
		
		\section{Fitting the dynamical hysteresis exponent}
		
		We follow the method used in ref. \cite{Bar_2018}.
		\subsection{Method 1: Best straight line fits on a log-log plot}
		The scaling exponent $\gamma$ was computed from fitting the area $A_i$ of the hysteresis loop (the transition temperatures $T_i$) observed for different temperature ramp rates R described by the following equation.
		
		\begin{equation}
			\begin{split}
				A_i=A_0 + \alpha R_i^{\gamma_a}\\
				T_i=T_0 \pm k R_i^{\gamma_t} 
			\end{split}
		\end{equation}
		\renewcommand\thefigure{3 (supplement)}
		\begin{figure}[h!]
			\includegraphics[scale=0.26]{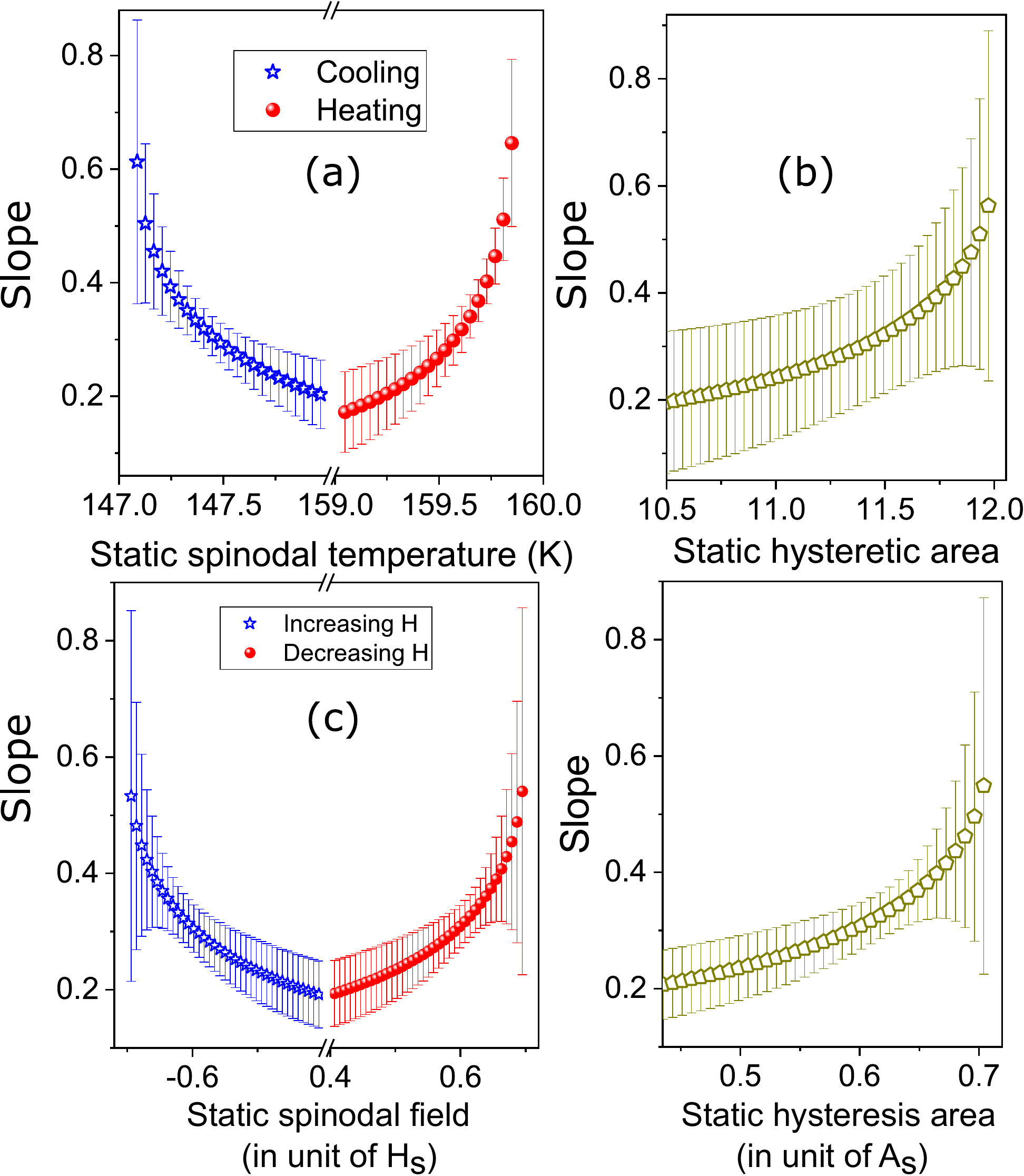}
			\centering
			\caption{(a) Slope versus heating and cooling quasistatic transition temperatures. (b) Slope versus quasistatic hysteresis area. The error in slope for each data point is also shown with error bars. We choose the model with noise strength 0.02. (c) Slope versus quasistatic transition fields during increasing field and decreasing field. (d) Slope versus quasistatic hysteresis area.}
			\label{Figure:slope_field}
		\end{figure}
		where, $A_0$ is the quasistatic $(R\to0)$ area of the hysteresis loop and $T_0$ the quasistatic cooling (heating) transition temperature for fitting the data during the cooling (heating) cycles. $\alpha$ and $k$ are unknown constants. The plus and minus signs correspond to heating and cooling respectively. The hysteresis area and the heating transition temperature increase with the increase of sweep rate while the cooling transition temperature decreases with the sweep rate increase. In the main text, we plot the $|T_i-T_0|$ and $|A_i-A_0|$ versus sweep rate ($R$) in log-log scale. The slopes of the straight lines denote the scaling exponent ($\gamma_a$ for area and $\gamma_t$ for transition temperatures). The estimation of the scaling exponent from the area and the transition temperatures are mutually independent but can be found using the same method. Here we describe the method of estimation of the scaling exponent from the area of hysteresis loop for different sweep rates thoroughly. A slight change in quasistatic area leads to large change in scaling exponent. We estimated the scaling exponent and corresponding error for different values of $A_0$ from its acceptable region. We choose the value of $A_0$ that gives the minimum error in the slope. The slope corresponding minimum error is the scaling exponent. Similarly, we can estimate the scaling exponent for heating and cooling transition temperatures \cite{Bar_2018,Bar_2021}.
		
		In Fig. \ref{Figure:slope_field} (a, b), we show that the slope is extremely sensitive to the inferred values of the quasistatic transition temperatures and area. Error is minimum at 147.44 K for cooling, 159.69 K for heating and 11.37 for area. In Fig. \ref{Figure:slope_field} (c, d), we show the same for the theoretical model. Error is minimum at $-0.598$ for decreasing field, $+0.606$ for increasing field and $0.62$ for area.
		
		In Fig. \ref{Figure:slope_field_error}, we show the error in slope with slope for different values of heating and cooling quasistatic transition temperatures (a) and quasistatic area (b). The error is minimum at $\gamma=0.31$ for cooling, $\gamma=0.36$ for heating and  $\gamma=0.3$ for hysteresis area. In Fig. \ref{Figure:slope_field_error} (c, d), we show the same for theoretical model with noise strength 0.02.
		
		\renewcommand\thefigure{4 (supplement)}
		\begin{figure}[h!]
			\includegraphics[scale=0.23]{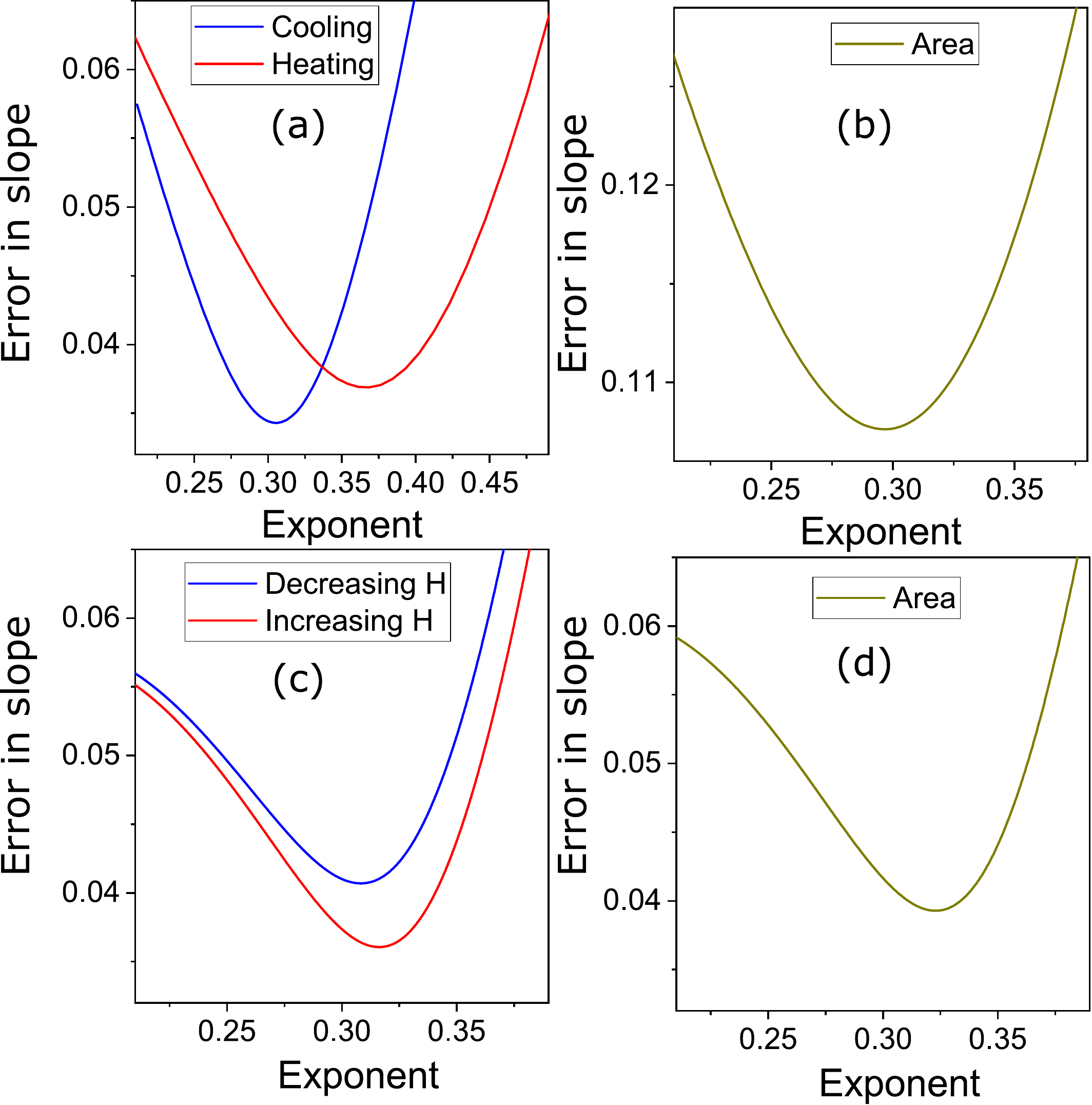}
			\centering
			\caption{(a) Error in the slope (discussed in previous figure) versus the slope on fitting Eq. \ref{Figure:power_law} to a straight line on log-log scale for different values of heating and cooling quasistatic transition temperatures. (b) Error in the slope versus the slope for different values of quasistatic hysteresis area. Error is minimum at $\gamma=0.31$ for cooling, $\gamma=0.36$ for heating and  $\gamma=0.3$ for hysteresis area. For comparison with experiment, we choose a model system with noise strength 0.02. (c) Error in the slope versus the slope for different values of quasistatic transition fields during increasing field and decreasing field. (d) Error in the slope versus the slope for different values of quasistatic hysteretic area.}
			\label{Figure:slope_field_error}
		\end{figure}

		\subsection{Method 2: Nonlinear fitting treating data points as independent quadruples} 
		
		We performed our experiment for 18 different sweep rates spunning almost two orders of magnitude. If $A_i$ is the area under the experimentally measured hysteresis curve (in T-S plane) at the sweep rate $R_i$, we assume $A_i$ increases with the sweep rate $R_i$ $[i = 1, .., 18]$, satisfying power law behaviour, i.e.,    
		
		\begin{equation}
			A_i=A_0 + \alpha R_i^\gamma
			\label{Equ:power_law_area}
		\end{equation}
		where $A_0$ and $\alpha$ is constant for this system. $A_0$ is the quasistatic $(R\to0)$ area of the hysteresis loop. Similarly, for another set of sweep rate ($R_j$) and corresponding area under hysteresis curve ($A_j$), we can write
		
		\begin{equation}
			A_j=A_0 + \alpha R_j^\gamma
			\label{Figure:power_law}
		\end{equation}
		We can eliminate the quasistatic area ($A_0$) term by subtracting one equation from another.
		\begin{equation}
			A_i-A_j = \alpha (R_i^\gamma -R_j^\gamma)
			\label{Figure:power_law_2}
		\end{equation}
		
		We can choose two points from total number of data points (say N) in a $^NC_2$ (say M) possible ways. We can remove the constant term $\alpha$ by dividing the Eq. 10 with the similar equation for another set $\{m,n\}$, i.e.
		
		\begin{equation}
			\frac{A_i-A_j}{A_m-A_n}=\frac{R_i^\gamma-R_j^\gamma}{R_m^\gamma-R_n^\gamma}
			\label{Figure:power_law_4}
		\end{equation}
		
		$\gamma$ is only unknown parameter in this above equation. We can solve the above transcendental equation to
		get value of $\gamma$. Overall we have $^MC_4$ transcendental equation to get as many  value of $\gamma$.

		\renewcommand\thefigure{5 (supplement)}
		\begin{figure}[h!]
			\includegraphics[scale=0.21]{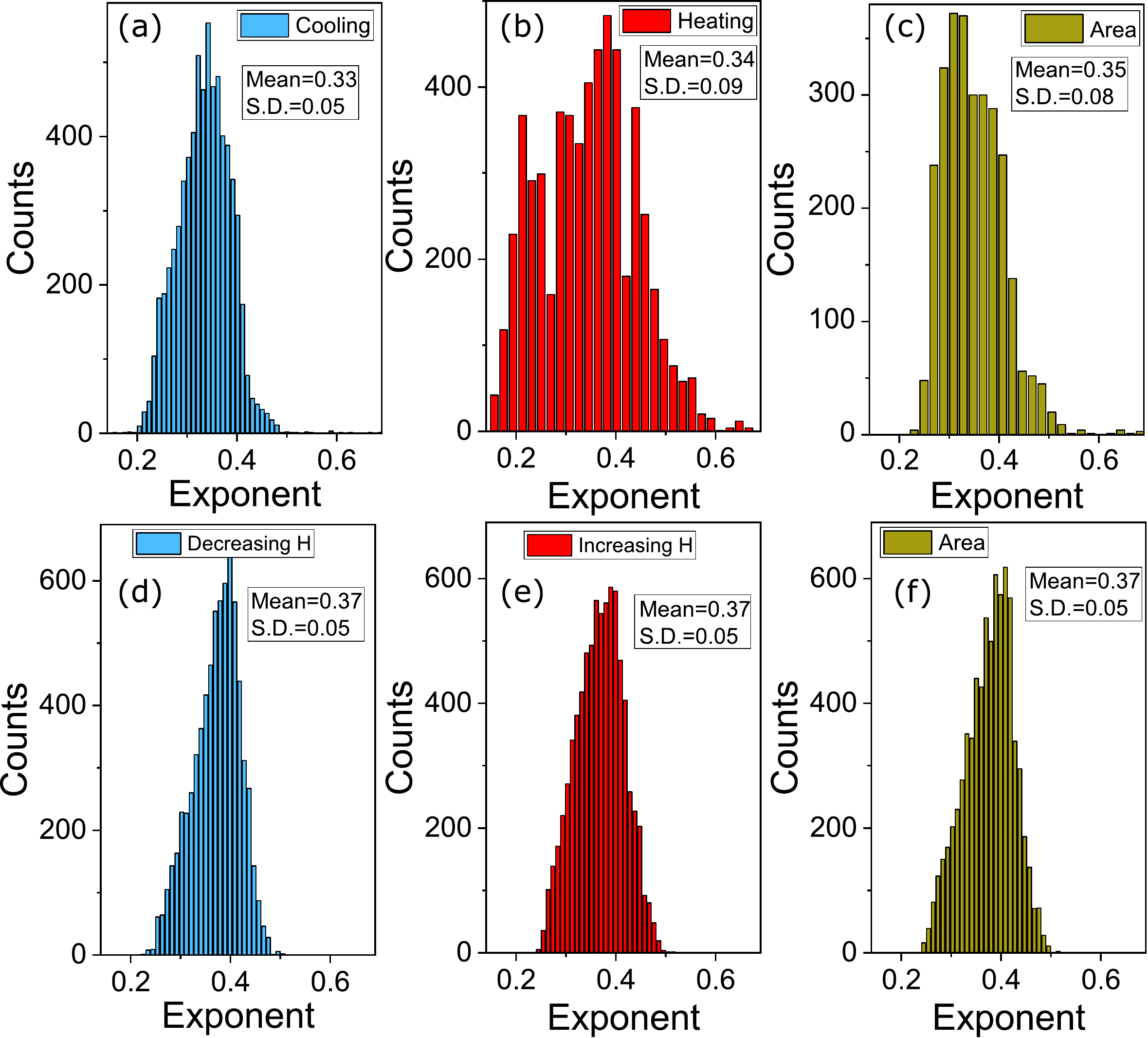}
			\centering
			\caption{Histogram method for estimating scaling exponent in an alternative manner. The histogram of scaling exponents calculated from feasible sets (satisfying Eq. \ref{Figure:power_law_4}) of cooling transition temperature (a), heating transition temperature (b) and area of the hysteresis (c) in the conjugate plane. For comparison with the theoretical model described previously, we show the same for the transition field associated with decreasing field (d), increasing field (e) and area of the hysteresis loop (f). Here we choose the noise strength ($\sigma$) of the system as 0.02 which nearly matches the experimental result shown in Fig. 8 (main text).}
			\label{Figure:histogram}
		\end{figure}
		
		
		For each computed $\gamma$ from the pairs (\{i, j\}, \{m, n\}), we get two values of $A_0$ from Eq. \ref{Figure:power_law} and Eq. \ref{Figure:power_law_2}, i.e.,

		\begin{equation}
			\begin{split}
				A_{0}^{\{i,j\}|\{m,n\}} &= \frac{A_i-\left(\frac{R_i}{R_j}\right) ^{\gamma}A_j}{1-\left(\frac{R_i}{R_j}\right) ^{\gamma}} \\
				A_{0}^{\{m,n\}|\{i,j\}} &= \frac{A_m-\left(\frac{R_m}{R_n}\right) ^{\gamma}A_n}{1-\left(\frac{R_m}{R_n}\right) ^{\gamma}} 
			\end{split}
			\label{Figure:sol_A0}
		\end{equation}

		
		These values can be plotted in a histogram and one may estimate the values of the mean and the dispersion around the mean for the estimates of $A_0$ and $\gamma$ [Fig. \ref{Figure:histogram}].
		
		In our experimental and numerical data, the area of hysteresis curve increases with the increase in sweep rate [Eq. \ref{Equ:power_law_area}], since Eq. \ref{Equ:power_law_area} implies that the quasistatic area ($A_{0}$) should be less than the area of hysteresis curve for lowest sweep rate (0.2 K/min). Furthermore, the difference between hysteresis area for the lowest sweep rate and the quasistatic hysteresis area should be less than the difference between the hysteresis area for high sweep rate ($R_1$) and hysteresis area for lowest sweep rate ($A_{0.2}$), i.e., $A_{0.2}-A_{0}<A_{R_1}-A_{0.2}$. Therefore, $A_0$ is bounded in region $(A_{0.2}-\delta A)<A_0<A_{0.2}$ where $\delta A$ be the measured value of hysteresis area difference ($|A_{R_1}-A_{0.2}|$) for two sweep rates (Let say $R_1=20$ K/min).
		
		Similarly, we can estimate the acceptable values for the scaling exponent from all pairs of transition temperatures (both cooling and heating) and corresponding sweep rates.

		
		In Fig. \ref{Figure:histogram} (a, b, c), we plot the histograms of the scaling exponents associated with the cooling transition temperature, the heating transition temperature and the area of the hysteresis curve. Bin number of the histogram is choosen in a way proposed by D.W. Scott \cite{Scott_1979}.
		
		\begin{equation}
			w_N=3.49\sigma N^{2/3}
		\end{equation} 
		where $\sigma$ is the standard deviation and N is the length of the data. We can approximate bin number by $b_N$=(max(data)-min(data))/$w_N$.
		
		The mean values [$\gamma=0.33$ (cooling), $\gamma=0.34$ (heating) and $\gamma=0.35$ (area)] estimated by this method are in excellent agreement with the best fit estimate that we have calculated in the previous section. The standard deviation in the histograms are 0.05 (cooling), 0.09 (heating) and 0.08 (area) and can be used as estimates of error. Note that these errors are of the same order of magnitude with the errors in slope mentioned in the previous method. Hence the values of the exponents are $\gamma=0.33\pm0.05$ (cooling) and $\gamma=0.34\pm0.09$ (heating) and $\gamma=0.35\pm0.08$ (area).
		
		In Fig. \ref{Figure:histogram} (d, e, f), we show the histogram of the scaling exponent for theoretical hysteresis (discussed in main text) of noisy systems (with noise strength 0.02).

	\end{document}